\documentclass[12pt,preprint]{aastex}

\shorttitle{External Compton Emission in Blazar Flares}
\shortauthors{Sokolov \& Marscher}

\received{}
\begin{document}

\title{External Compton Radiation from Rapid
Nonthermal Flares in Blazars}

\author{Andrei Sokolov\altaffilmark{1} and Alan P. Marscher} 
\affil{Institute for Astrophysical Research, Boston
University, 725 Commonwealth Avenue, Boston, MA 02215,
USA}
\altaffiltext{1}{present address: Center for Astrophysics,
University of Central Lancashire, Preston PR1~2HE, UK}
\email{asokolov@uclan.ac.uk, marscher@bu.edu}

\begin{abstract}
In this paper we extend our approach to modeling
multifrequency
emission variability from blazars to include
external-radiation Compton
(ERC) emission and electron energy losses from inverse
Compton
scattering of seed photons originating outside the
jet.
We consider seed-photon emission from a dusty
molecular torus and the 
broad line region (BLR) surrounding the
central engine.
We establish constraints on the properties of
the molecular torus and BLR under which synchrotron
self-Compton (SSC) 
emission dominates such that the results obtained in
our previous paper are 
applicable. The focus of this study is on relative
time delays between
the light curves observed at different frequencies.
For definiteness,
we consider emission resulting from a collision
between relativistic 
shocks, but the results apply more generally to
conditions involving 
acceleration of relativistic electrons and/or magnetic
field 
amplification at any type of front. Unlike SSC
emission, ERC flares 
involving a constant field of seed photons are not
delayed by light 
travel time of the seed photons. The main cause of
delays is from 
radiative energy losses, which result in frequency
stratification 
behind the front and quench the flare first at the
highest frequencies, 
progressing to lower frequencies as time advances.
However, if the spectrum of electrons injected at the
shock front
is characterized by a relatively high value of the
minimum energy
(a Lorentz factor $\gamma_{{{}} min}\sim100$ is
sufficient),
the ERC flare in the X-ray band can be delayed and
may even peak after 
the injection has ceased. This effect is strongly
frequency dependent, 
with a longer lag at lower frequencies and an X-ray
spectral index that 
changes rapidly from positive (inverted spectrum) to
steep 
values. 
\end{abstract}


\section{Introduction}
In \citet*[hereafter Paper I]{paper1}, we 
introduced a model of blazar variability
designed to study rapid flares resulting from particle
acceleration at a shock front.
The case considered involved a collision between
relativistic shocks,
but the effects incorporated, light-travel delays and
energy/frequency
stratification, are of general importance. 
The primary goal of that study was to establish the
conditions
for relative delays between synchrotron and
synchrotron self-Compton (SSC) flares at different
frequencies.
We found that the SSC emission can be strongly
affected
by light travel time of the synchrotron seed photons,
which can result in a considerable delay of the SSC
flares in the X-ray 
band with respect to synchrotron variability at lower
frequencies.
In the present paper, we consider external sources of
the seed photons
that are scattered to high energies by the
relativistic electrons
heated at the front.

The current unified scheme of active galactic nuclei \citep{ant93}
includes a source of obscuration that shields the
emission
from the region within $\sim$ 0.1-10 pc of the central
black hole
when observed at large viewing angles.
This obscuration is usually assumed to be provided by
a dusty
molecular torus that surrounds the innermost nuclear
region.
The dust in the inner portions of the torus facing the
central
continuum source is heated to temperatures
$T\sim1000\,\mbox{K}$
and emits infrared radiation roughly as a blackbody.
In addition, reverberation studies of broad
emission lines 
\citep{pet93} suggest
the presence of a clumpy broad emission line region
(BLR)
that consists of ionized clouds concentrated within
$\sim0.1L_{{{}} uv,42}^{1/2}\,\mbox{pc}$ of the black
hole,
where $L_{{{}} uv,42}$ is the UV spectral luminosity
normalized by 
$10^{42}\,\mbox{erg}\,\mbox{s}^{-1}\,\mbox{\AA}^{-1}$.
The combined BLR  emission and blackbody
radiation from the torus
permeate the jet on parsec scales, providing a
potentially
significant source of seed photons,
alongside the synchrotron seed photons produced
by the jet plasma itself \citep[e.g., see][]{bla00}.

In the rest frame of the emitting plasma, 
radiation from the torus and BLR  depends
on
location of the emission region in the jet and 
its
bulk Lorentz factor $\Gamma'_p$.
Radiation from the torus is in general expected
to be Doppler boosted in the plasma rest frame. This
results in 
stronger dependence of the ERC emission
on $\Gamma'_p$ compared with the SSC emission, whose
seed 
radiation is produced in the rest frame of the plasma.
However,
this behavior can be modified by the effects of
electron energy losses 
from ERC radiation which, if dominant, limit the
spatial
extent of the emitting plasma in an energy-dependent
manner, leading
to frequency stratification of the emission.

In this paper, we investigate the broadband features
of the 
ERC emission during flares under different assumptions
about 1) properties of the molecular torus, 
2) location in the jet where the flare occurs, and
3) the value of $\Gamma'_p$ of the emitting plasma.
As in Paper~I, we concentrate on the study of
relative
delays between flares at different frequencies and the
features
that can be used to distinguish among different
emission mechanisms.
An additional goal of this study is to define the
properties of the molecular torus and the BLR under
which
the results obtained in Paper~I are valid.
There
we assumed that the ERC emission provides a negligible
contribution 
compared with the SSC radiation at the frequencies of
interest,
and that the energy losses of electrons are dominated by
synchrotron 
emission.

\section{External Compton Model}
The inclusion of ERC radiation in our model of
variability
introduces a new component of high-energy emission.
It can also change the  overall structure of the
emission region
and, hence, affect synchrotron and SSC radiation if 
the energy losses of electrons are dominated by
scattering of
external photons.
In this case the decay time of electrons will be
reduced 
compared with the case of pure synchrotron losses,
which leads to changes in flux levels and values
of critical frequencies, such as the break frequency.
The details depend on the structure and properties of
the sources
of external emission, which we describe in this
section.

As in Paper~I, here we concentrate on the study
of rapid 
variability on time scales $\sim1\,\mbox{day}$.
We adopt the assumptions about geometry and excitation
structure of the emitting volume made in Paper~I.
We ignore the expansion of the emission region, which
is assumed to be a cylinder oriented along the jet.
We assume that the size of the emitting volume is small 
compared to the sizes of  and distances to the external
sources of emission.
Then we can assume that the external radiation is
homogeneous 
throughout the emitting plasma, albeit highly anisotropic.
The structure and location of the molecular torus and
the BLR,
as well as the location of the emitting plasma in the
jet,
determine the angular dependence of external emission.
This anisotropy is further amplified by relativistic
aberration
and Doppler boosting or de-boosting in the rest frame
of the plasma.
For simplicity we assume that external radiation is static during the flare.

The properties of the putative dusty torus in
the nucleus of an active galaxy are poorly known.
In particular, despite what its name implies, the
geometrical shape and 
size of this structure in a quasar or BL~Lac object
are poorly 
constrained. According to one model \citep{elv00}, the
obscuration
is provided by a conical outflow of the material from
the accretion 
disk. Recent interferometric observations of the
nucleus of NGC~1068 
\citep{jaf04} reveal the presence of warm dust at
temperature 
$T\sim300\,\mbox{K}$ in a structure
$\sim2.1\,\mbox{pc}$ in size,
surrounding a smaller, warmer ($T>800\,\mbox{K}$)
structure
of size $\sim0.7\,\mbox{pc}$.
The mass of the black hole in NGC~1068 is $1.4\times 10^7\mbox{M}_\sun$ 
according to VLBI measurements of water maser emission \citep{gri97}.
For quasars and BL~Lac objects harboring more massive black holes, one should 
expect the size of the torus to scale accordingly.

\begin{figure}
%\epsscale{0.6}
\plotone{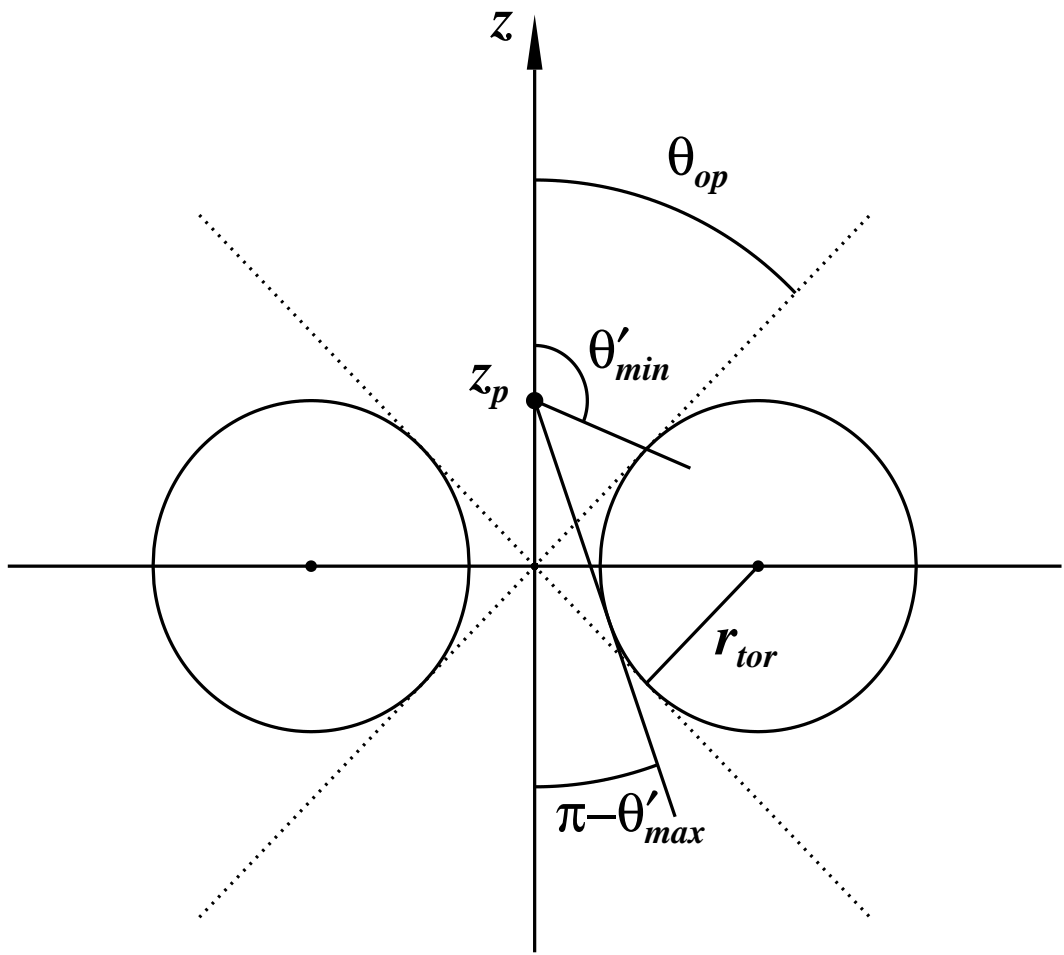}
\caption{
A schematic diagram of the central region of an active
galaxy
showing the location of the relativistic plasma in
the jet at position $z_{{{}} p}$ in relation to 
the molecular torus with semi-opening angle
$\theta_{{{}} op}$ and radius $r_{{{}} tor}$. In this
sketch,
$\theta_{{{}}
op}=45^\circ$, and
$r_{{{}} tor}=z_{{{}} p}$.
The incident emission from the torus is assumed to be
constrained 
between the minimum and maximum angles, $\theta'_{{{}}
min}$ and $\theta'_{{{}} max}$, respectively.
Only a fraction of the torus that faces the center
is assumed to be hot enough to contribute substantially
to the seed photon field at $z_p$.
\label{ext}}
\end{figure}

Fig.~\ref{ext} illustrates the geometry and size 
of the molecular torus in relation to
the position of the emitting blob of plasma in the jet
for a representative set of assumptions about
the external sources of seed emission and the
location of the radiating plasma.
The torus is characterized by semi-opening angle
$\theta_{{{}} op}$
and radius $r_{{{}} tor}$.
We assume that the emission from the torus is
dominated by
dust that radiates as a black body at temperature $T$,
so that the intensity of emission from the torus is
\begin{equation}
I'_{\nu'}(\theta')={\cal %script 
B}_{\nu'}(T),\quad\mbox{for}\quad\theta'_{{{}}
min}<\theta'<\theta'_{{{}} 
max},
\end{equation}
where ${\cal %script 
B}_{\nu'}(T)$ is the Planck function. Here and below, the
primed quantities 
associated with the emitting plasma are given in the
rest frame of the 
host galaxy, whereas unprimed quantities are reserved
for use in the 
plasma rest frame (this convention follows the one
adopted in Paper~I). 
We only take into account the emission from the
portion of the torus 
that faces the central continuum source.
However, the details of this approach are not crucial
to
the final results.
The only essential parameters are the angle $\theta'_{{{}} min}$, 
which determines the maximum Doppler boosting,
and the dust temperature $T$.

The BLR can be represented by a uniform spherical source
of 
emission at a fiducial frequency $\nu'_{{{}}
blr}$. The integrated 
intensity of the incident emission from the BLR is
given by
\begin{equation}
I'(\theta')=\frac{L_{{{}}
blr}/(4\pi)}{\frac{4}{3}\pi{}r_{{{}} 
blr}^3}\Delta{r}(\theta'),\quad\mbox{for}\quad
\theta'_{{{}} 
blr}<\theta'<180^{\circ},
\end{equation}
where $r_{{{}} blr}$  is the radius of the BLR, 
$\Delta{r}(\theta')$ is the geometric thickness of the
BLR in a given 
direction, $\theta'_{{{}} blr}=\pi-\arcsin(r_{{{}}
blr}/z_{{{}} p})$,
and $L_{{{}} blr}$ is the BLR luminosity, which can be
estimated from 
observations if the distance to the blazar is
known.
Again, as in the case of the molecular torus,
the angle $\theta'_{{{}} blr}$ and the BLR luminosity $L_{{{}}
blr}$ are the only 
essential parameters.

The UV spectral luminosity of 3C~273 can been estimated as $L_{{{{}} uv},42}\approx2$ \citep{von97},
which gives the size of the BLR region as $\sim0.14\,\mbox{pc}$.
Although the size of the torus in 3C~273 is unknown, it is reasonable to expect
that it is at least $10\,\mbox{pc}$. 
For the calculations reported in this paper, we assume that the location of the emitting plasma
$z_{{{}} p}\approx r_{{{}} tor}$.
Under this condition the contribution of the BLR region to the seed photon field
is negligible since BLR emission is significantly de-boosted by relativistic Doppler
effects.
We therefore neglected the contribution from the BLR in the reported results.
However, the role of the BLR would increase sharply
if the event that causes the flare were to occur
closer to the central 
engine and, hence, closer to or even within the BLR.

The plasma  that produces variable emission during a
flare
moves down the jet at a relativistic speed
$v'_p =c\beta'_p$, with corresponding
Lorentz factor $\Gamma'_p$.
The Doppler effect and relativistic abberation cause the intensity of external radiation to be 
highly anisotropic in the plasma rest frame.
The direction of incoming radiation is modified
by relativistic aberration according to 
\begin{equation}
\mu=(\mu'-\beta'_p)/(1-\beta'_p\mu'),
\end{equation}
where $\mu'=-\cos{\theta'}$ and  $\theta'$ is the angle
of propagation of the incident photons relative to the
line of sight; $\mu$ is the 
corresponding quantity in the plasma rest frame.
The spectral intensity of incident emission from the
molecular torus is
transformed according to $I_{{{}}
\nu}(\mu)=\delta^3I'_{{{}} 
\nu'}(\mu')$, where the Doppler factor
$\delta=\Gamma'_p(1-
\beta'_{{{}} p}\mu')$ and $\nu=\delta\nu'$,
while the expression for transforming incident
integral
intensity is $I(\mu)=\delta^4I'(\mu')$.
Thus, for the parameters used in Fig.~\ref{ext},
the maximum Doppler factor for the emission from the
torus 
$\delta_{{{}} max}\approx\Gamma'_p$.

\begin{figure}
%\epsscale{1.0}
\plotone{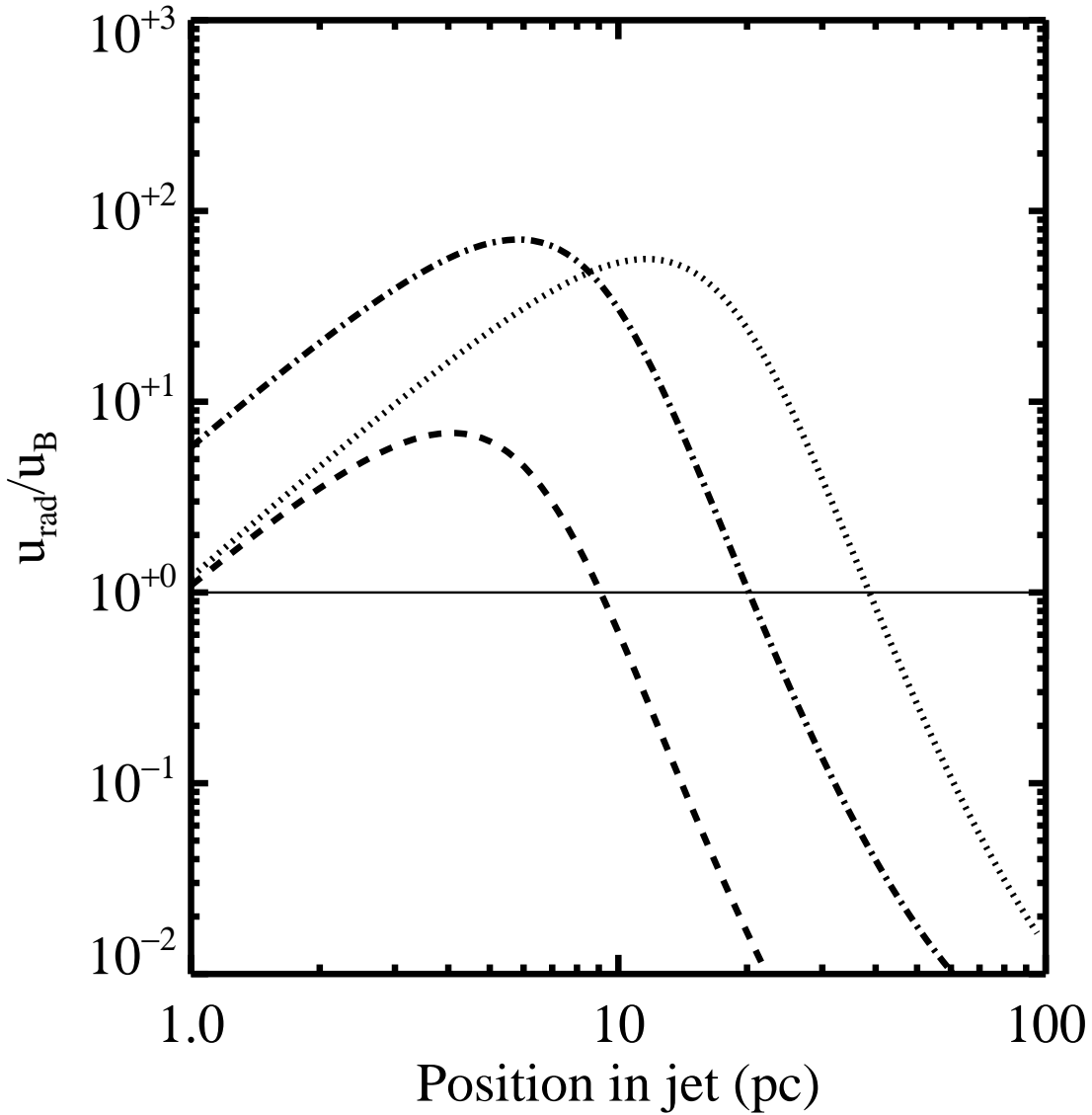}
\caption{
Ratio of the energy density of the external radiation field
to that of the magnetic field for three models of the molecular
torus,
as a function of position of emitting plasma along the jet $z_p$,
in the frame of the jet plasma for bulk Lorentz factor ${\Gamma'}_p=10$.
The semi-opening angle of the tori is $45^{\circ}$ for all three models.
The plot compares contributions from (1) moderately
hot dust, 
$T=800\,\mbox{K}$, within a torus of smaller size,
$r=7.0\,\mbox{pc}$ 
(long-dashed curve), (2) the same but with larger
size, 
$r=20\,\mbox{pc}$ (dotted curve), and (3) a hotter
torus with 
$T=1200\,\mbox{K}$ and $r=10\,\mbox{pc}$ (dot-dashed
curve).
The adopted rest-frame value of the magnetic field 
strength is $B=B_0z_0/z_p$, with $B_0=0.4\mbox{G}$ and $z_0=10\mbox{pc}$.
\label{dust.all}}
\end{figure}

Once the intensity as a function of direction is
known,
one can determine the value of the external radiation
energy density in the rest frame of the emitting
plasma
by integrating over all allowed incident directions
and over the black-body spectrum.
The external radiation energy density from three different models
of the molecular torus is presented in Fig.~\ref{dust.all}
as a function of position of the emitting plasma in
the jet.
The position of the emitting plasma is a free parameter 
of the model. It does not change during the calculations,
but adopting a different value may affect the results of the simulations considerably.
The energy density of the magnetic field in the plasma 
rest frame  $u_B=B^2/(8\pi)$,
where the magnetic field strength as a function of position $z_p$
along the jet is given by  $B(z_p)=B_0z_0/z_p$.
It can be seen that the energy density of
emission from the 
torus exceeds that of the magnetic field for
$z_p\lesssim40\,\mbox{pc}$ when 
the magnetic field $B_0$ is within half an order of
magnitude of our
adopted value of 0.4~G. 

The expression for the energy density of blackbody
radiation in the rest 
frame of the plasma can be integrated to produce the
following 
approximate expression:
\begin{equation}
\label{uradapprox}
u_{{{}}
rad}\approx\frac{2\pi}{c}\sigma{}T^4\frac{\delta^3_{{{}}
max}}{3\Gamma'_p},
\end{equation}
where $\sigma$ is the Stefan-Boltzmann constant.
When the location of emitting plasma $z_p\approx{}r_{tor}$, one has $\theta'_{min}\approx90^{\circ}$
and the corresponding $\mu'\approx0.$
Under these conditions the ratio of the external radiation and magnetic energy densities
 has a
simple dependence on the physical parameters:
\begin{equation}
u_{rad}/u_B\approx 30 \left(\frac{T}{1200\,\mbox{K}}\right)^4
\left(\frac{\Gamma'_p}{10}\right)^2
\left(\frac{B}{0.4\,\mbox{G}}\right)^{-2}.
\end{equation}
It can be seen that moderate changes
in the parameters can result in the case where magnetic energy density 
dominates and, therefore, energy losses of electrons are primarily due to synchrotron (or SSC) emission.
For $T=800\mbox{K}$ and $B=1\mbox{G}$, $u_{rad}\approx{}u_B$ if $\theta'_{min}\approx90^{\circ}$.
Unfortunately, no simple approximation is available for the dependence of $u_{rad}$ on
the location of the emitting plasma $z_p$ when it is 
different from $z_p\approx{}r_{tor}$.
If the size of the torus is considerably larger than $z_p$,
the angle $\theta'_{min}\approx\theta_{op}$, which generally results
in a somewhat larger value of $\delta_{max}$ and, hence,
 even more pronounced dominance of $u_{rad}$ over $u_B$.
On the other hand, if $z_p>r_{tor}$, the contribution from the torus
is diminished rapidly with increasing $z_p$.
Since $u_{{{}} rad}$ can exceed $u_B$ when $z_p \lesssim{}r_{tor}$,
it must in general be taken into account when
considering 
electron energy losses, which can be expressed as
$\dot\gamma=-\gamma^2/t_{{{}} u}$,
where
\begin{equation}
t_{{{}} u}=\frac{7.73\times10^8\,\mbox{s}}{8\pi(u_{B}+u_{{{}} rad})}.
\end{equation}
The formalism developed in Paper~I
can be recovered completely by substituting $t_1$
defined there
with $t_{{{}} u}$.
The ERC flux $F^E_{{{}} \nu}(t_{{{}} obs})$ as a
function of time
$t_{{{}} obs}$ and frequency $\nu$ of observation can
then be calculated
by utilizing the same procedures that are employed in the SSC
calculations.

\section{Calculated Spectra and Light Curves}
\begin{figure}
\epsscale{0.6}
\plotone{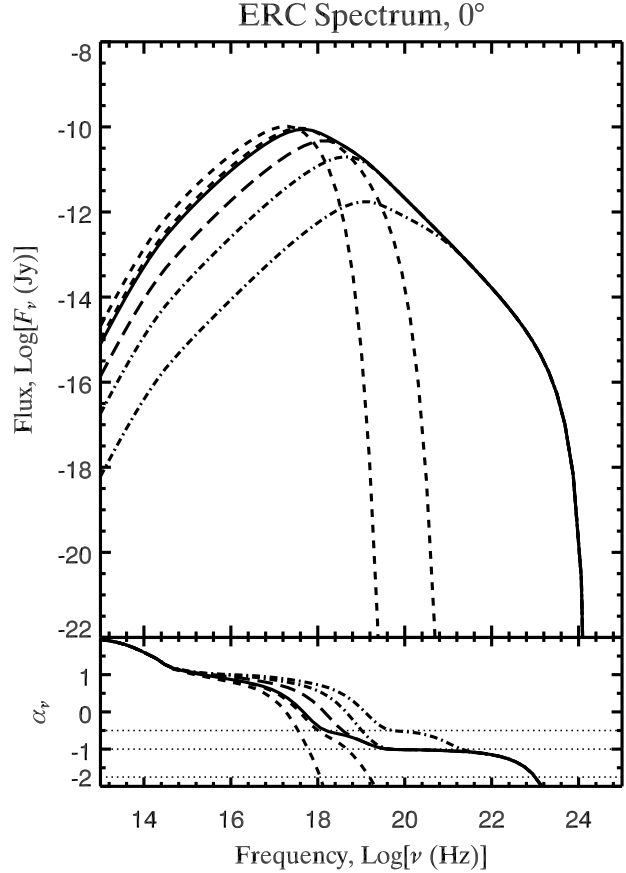}
\caption{
The evolution of the broadband external Compton
spectrum
for viewing angle $\theta_{{{}} obs}=0^{\circ}$. The
calculation for
this and subsequent figures is carried out for an
observer at rest
relative to the emitting plasma.
Only the contribution from  the forward-shock region
is considered.
The individual spectra shown  are for the following
times $t/t_{{{}} ac}$:
0.02, 0.2 (dot-dashed), 0.5 (long dashed), 1.0
(solid), 1.1, 1.5 (short dashed).
The bottom panel indicates the evolution of the 
spectral index $\alpha_{{{}} \nu}$, 
defined by $F_{\nu}(t_{{{}}
obs})\propto\nu^{\alpha_{\nu}}$.
The following input parameters describing 
the shocked plasma were used:
$R=H=4\times10^6\,\mbox{s}$,
$v=0.34c$, $B=0.4\,\mbox{G}$, $n=10\,\mbox{cm}^{-3}$,
$\gamma_{{{}} min}=100$, $\gamma_{{{}}
max}=2\times10^4$, power-law 
electron energy distribution of slope $-s$ with $s=2$,
and fiducial distance $D_0=10^3\,\mbox{Mpc}$.
These are defined in the rest frame of the emitting
plasma and identical to the ones used in Paper~I.
The emitting plasma is assumed to be located at
$z_{{{}} p}=10\,\mbox{pc}$ in the jet
and moving downstream at Lorentz factor
$\Gamma'_p=10$.
The external emission is characterized by the
parameters used in Fig.~\ref{dust.all}, model (3).
\label{mesp.0}}
\end{figure}

\begin{figure}
%\epsscale{1.0}
\plotone{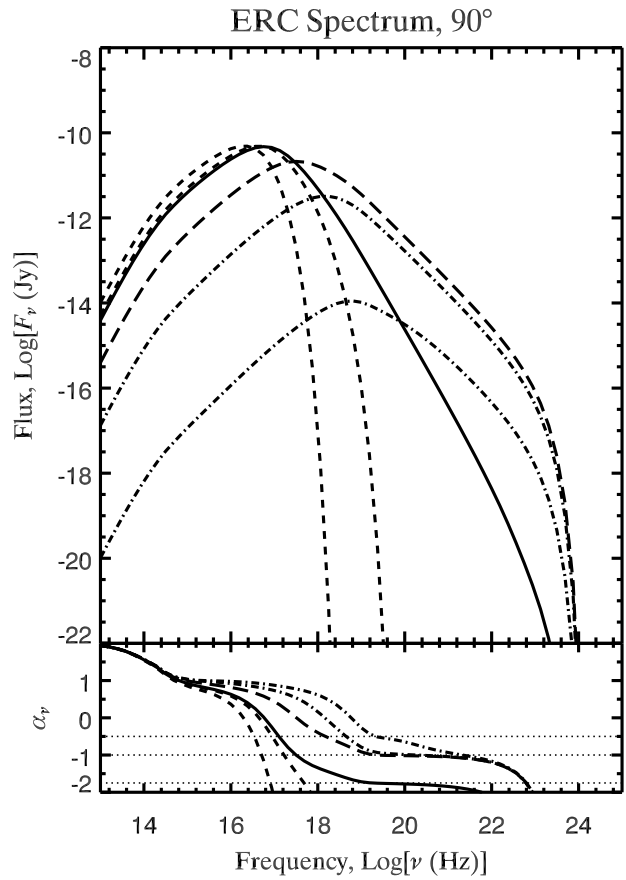}
\caption{
The same as in Fig.~\ref{mesp.0} but for a viewing
angle of 
$90^{\circ}$ in the rest frame of the emitting plasma,
or
$\sin^{-1}(1/\Gamma'_{{{}} p})$ in the observer's
frame.
\label{mesp.90}}
\end{figure}

\begin{figure}
%\epsscale{1.0}
\plotone{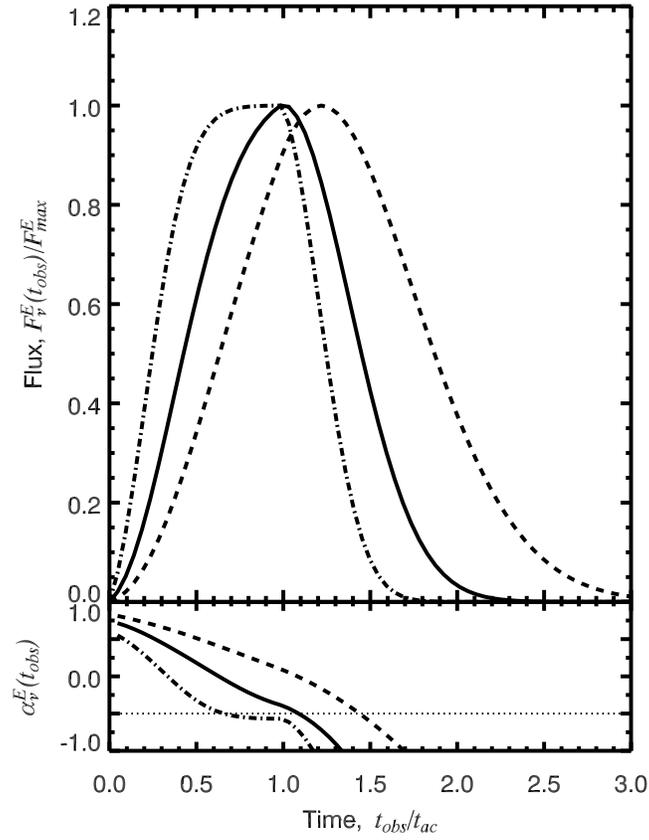}
\caption{
Normalized external Compton light curves for viewing
angle $\theta_{{{}} obs}=0^{\circ}$
at frequencies of $4\times10^{17}\,\mbox{Hz}$
(dashed), $10^{18}\,\mbox{Hz}$ (solid),
and $2.5\times10^{18}\,\mbox{Hz}$ (dot-dashed).
The time dependence of the spectral index for each
light curve is shown in the lower panel. 
\label{melc.0}}
\end{figure}

\begin{figure}
%\epsscale{1.0}
\plotone{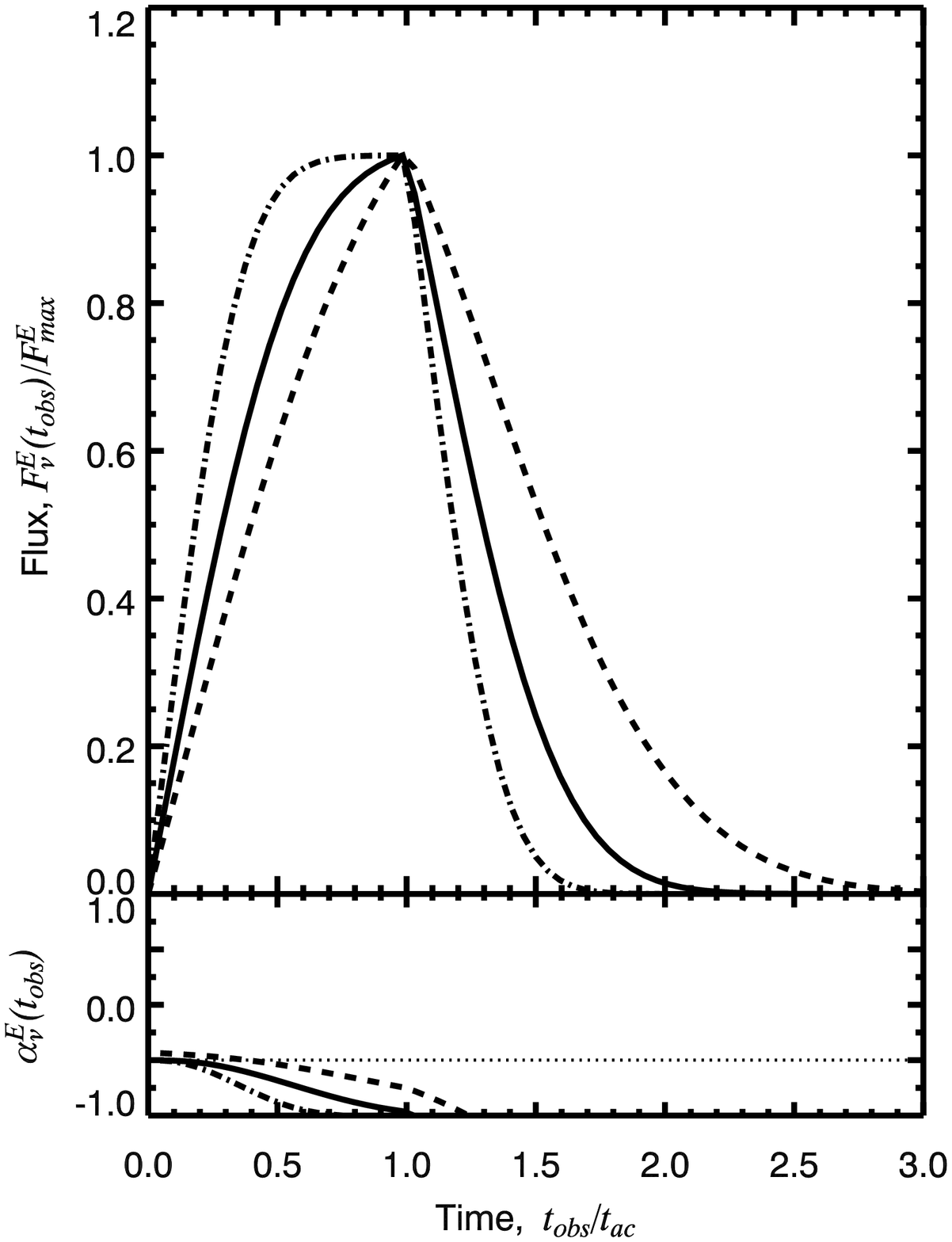}
\caption{
The same as Fig.~\ref{melc.0} but for $\gamma_{min}=10$.
\label{melc.0a}}
\end{figure}

\begin{figure}
%\epsscale{1.0}
\plotone{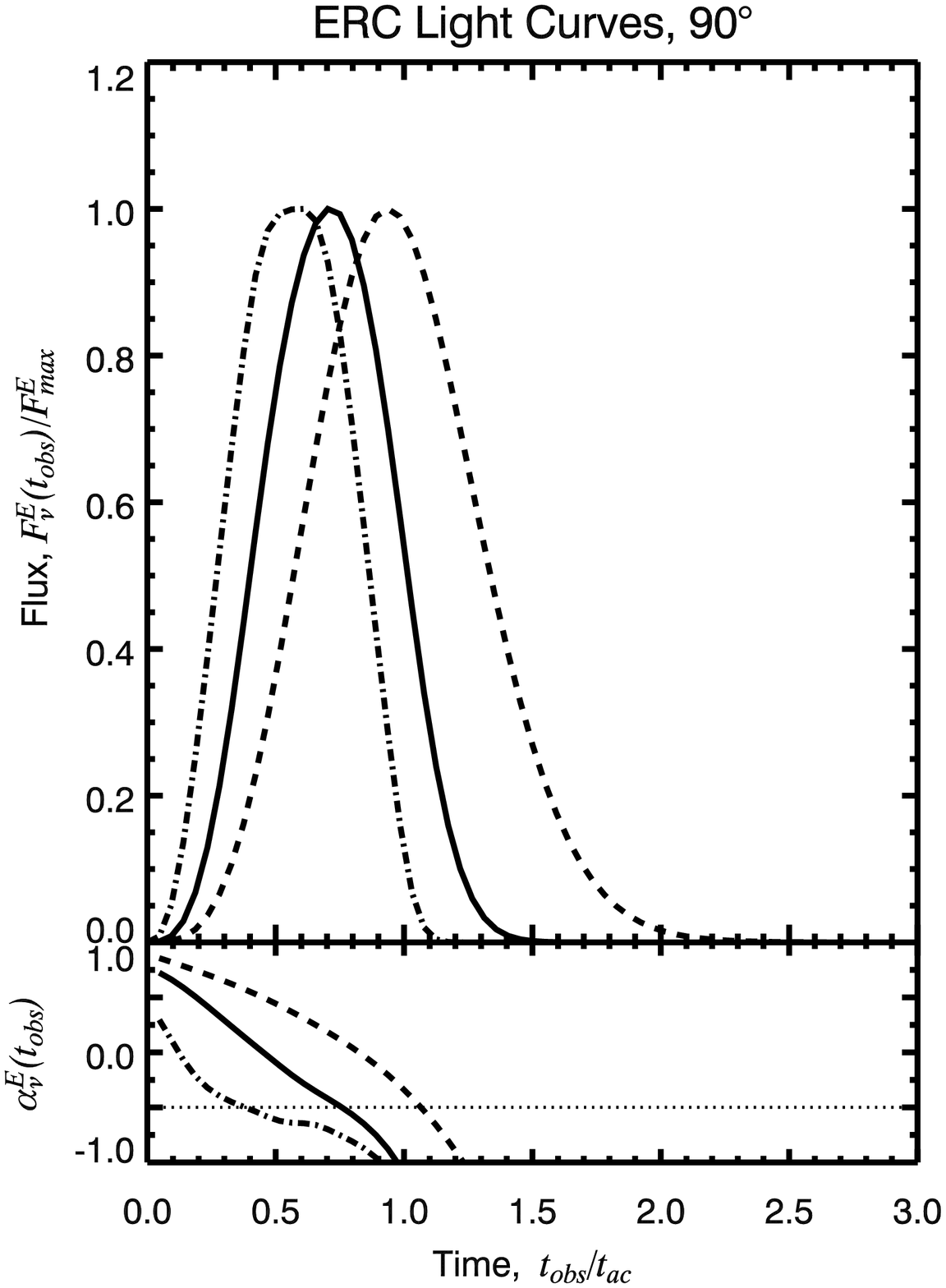}
\caption{
Normalized external Compton light curves for viewing
angle $\theta_{  obs}=90^{\circ}$, $\theta'_{obs}=\sin^{-1}(1/{\Gamma'}_{  p})$,
at frequencies of $10^{17}\,\mbox{Hz}$
(dashed), $4\times10^{17}\,\mbox{Hz}$ (solid),
and $2.5\times10^{18}\,\mbox{Hz}$ (dot-dashed).
The time dependence of the spectral index for each
light curve is shown in the lower panel. 
%The same as Fig.~\ref{melc.0} but for viewing
%angle $\theta_{{{}} obs}=90^{\circ}$, $\theta'_{{{}}
%obs}=
%\sin^{-1}(1/\Gamma'_{{{}} p})$.
\label{melc.90}}
\end{figure}

\begin{figure}
%\epsscale{1.0}
\plotone{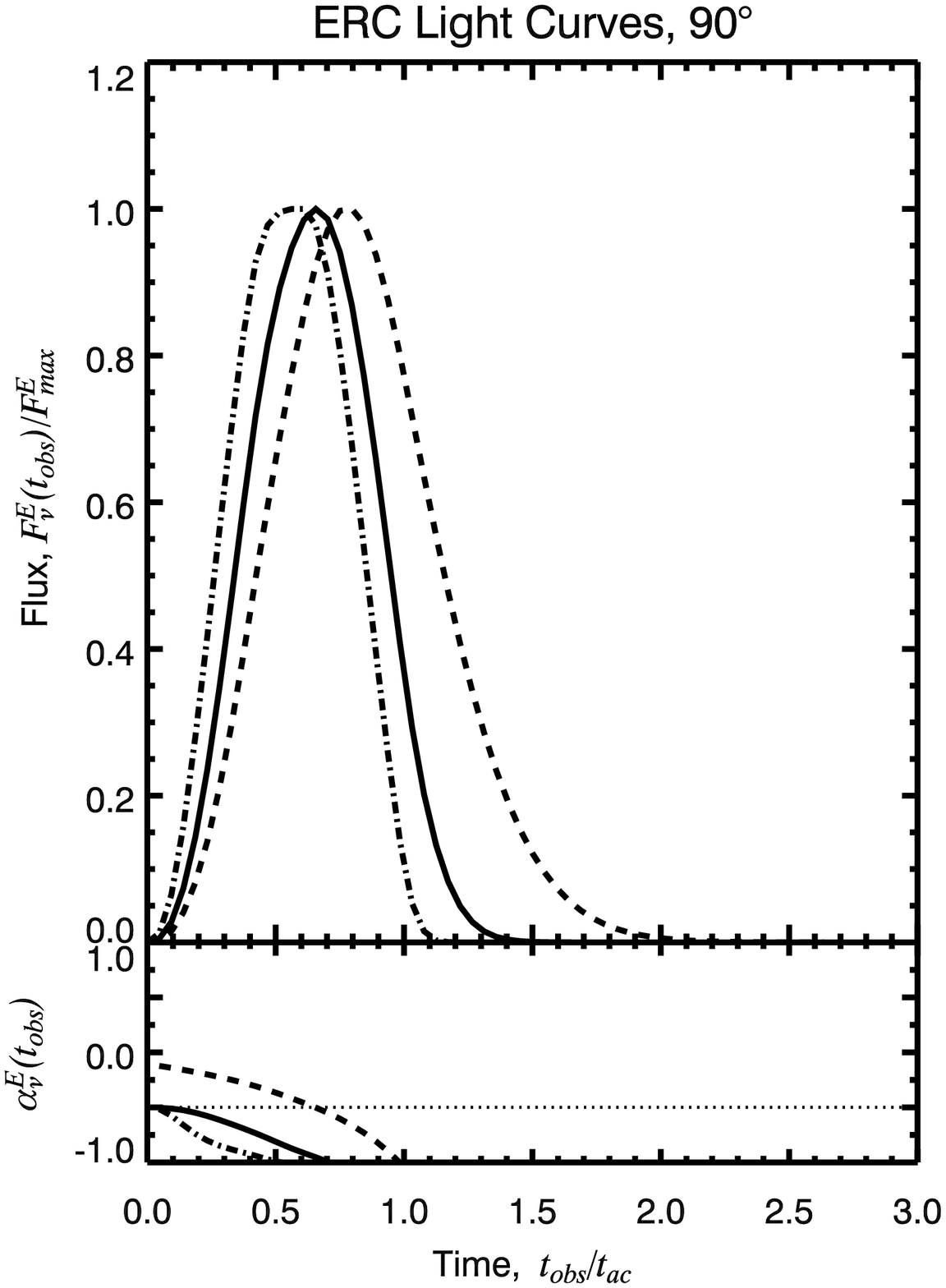}
\caption{
The same as Fig.~\ref{melc.90} but for $\gamma_{min}=10$.
\label{melc.90a}}
\end{figure}

In this section, we present the results of simulations
of ERC 
emission variability. We study both the broadband
spectral variability 
(Figs.~\ref{mesp.0} and~\ref{mesp.90}) and
the light curves at several representative frequencies
(Figs.~\ref{melc.0} through~\ref{melc.90a})
for two viewing angles: $\theta_{{{}} obs}=0^{\circ}$
corresponding to 
the case when the line of sight coincides with the jet
axis, and 
$\theta_{{{}} obs}=90^{\circ}$
in the rest frame of the plasma, which maximizes
superluminal motion
and reduces to the small angle $\theta'_{{{}}
obs}\sim1/\Gamma'_p$
in the frame of the host galaxy.
To characterize external emission we specify the bulk
Lorentz
factor of the emitting plasma   $\Gamma'_p=10$.
All the results presented in this section are given in
the rest frame 
of the emitting plasma.

\begin{figure}
%\epsscale{1.0}
\plotone{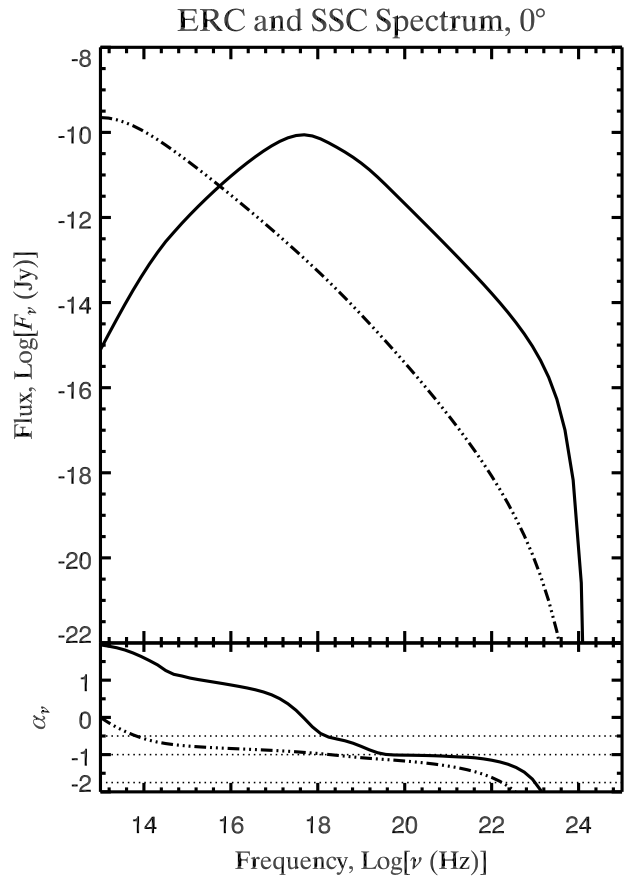}
\caption{
Synchrotron self-Compton (triple-dot-dashed curve) and
external Compton (solid curve) spectrum for
$\theta_{{{}} obs}=0^{\circ}$ at the crossing time, $t=t_{{{}} ac}$. 
The emitting plasma is described by the same input
parameters as in Fig.~\ref{mesp.0}, while the
parameters for the torus are 
the same as in Fig.~\ref{dust.all}, model (3).
ERC radiation dominates over SSC radiation at frequencies above
$10^{16}\,\mbox{Hz}$.
The SSC spectrum is different from the one presented
in Paper~I 
because the parameters have been chosen such that the
electron energy 
losses are dominated by ERC emission (see
Fig.~\ref{dust.all}).
\label{mesp3.0}}
\end{figure}

We use the same input parameters for the emitting
plasma as the ones 
employed in the companion SSC calculations. The
parameters describing 
external sources of emission, as well as the bulk speed
of the emitting 
plasma, are chosen such that the ERC flux dominates
over the SSC radiation at the frequencies of interest.
This corresponds to the external radiation energy
density in the plasma 
rest frame exceeding the energy density of the
magnetic field.
This affects the decay time of electrons and,
therefore,
the synchrotron and SSC emission variability.
The comparison between the SSC and ERC spectral energy distribution (SED)
at high frequencies is shown in Fig.~\ref{mesp3.0}.
Scattered photons from the molecular torus 
dominate at frequencies above $10^{16}\,\mbox{Hz}$ in
the calculations.
Scattered BLR radiation is not shown, since it provides a negligible 
fraction of the external photons because of severe
Doppler de-boosting. 

\subsection{Spectral Evolution}
Figs.~\ref{mesp.0} and~\ref{mesp.90} show a sequence
of SEDs from the forward-shock
region 
(see Paper~I) at different times normalized by
the apparent crossing 
time $t_{ac}$.
The latter quantity 
is defined in terms of the size of the
excitation zone,
which extends over a span $2R$ across and $H$ along
the jet, the speed 
of the shock in the plasma rest frame $v$,
and the viewing angle: $t_{{{}} ac}=[(c/v)-1]H/c$ for
$\theta_{{{}} obs}=0^{\circ}$
and $t_{{{}} ac}=2R/c+H/v$ for $\theta_{{{}}
obs}=90^{\circ}$ (for more details see Paper~I).
The spectral features of the ERC emission depend
on the
characteristic frequency of the infrared photons from
the torus 
at temperature $T$: $\nu_{{{}} tor}\sim 3kT/h$.
The ERC spectrum can be characterized by three
critical frequencies.
(1) The spectrum drops off exponentially above
frequency
$\nu_{{{}} e,max}(t=0)\approx 4\gamma^2_{{{}}
max}(t=0)\delta_{{{}} max}\nu_{{{}} tor}$
until $t=t_{{{}} ac}$, after which the drop-off
frequency begins to fall 
quickly due to the sharp decrease in the maximum value
of the Lorentz 
factor of electrons in the plasma owing to radiative
cooling.
(2) The turn-over frequency of the SED,
$\nu_{{{}} e,t}\approx4\gamma^2_{{{}} min}(t_{{{}} obs})
\delta_{{{}} max}\nu_{{{}} tor}$, decreases during the
flare  as $\gamma_{{{}} min}$ declines 
from the 
initial value, $\gamma_{{{}} min}(t=0)$.
 The value of the turn-over frequency at the
crossing time 
$t_{{{}} ac}$ depends on the energy loss rate of
electrons, which
in turn depends on the bulk speed of the plasma
if ERC losses are dominant.
(3) A break frequency $\nu_{{{}} e,b}$ is well defined
in the case of 
the ERC spectrum since the seed photon SED is nearly
monochromatic.
It is found by solving the equation
$t^E_{\nu}=\min\{t_{{{}} obs},t_{{{}} ac}\}$ for $\nu$,
where the decay time at frequency $\nu$ is defined as 
\begin{equation}
t^E_{\nu}=t_{{{}} u}\sqrt{\frac{4\delta_{{{}}
max}\nu_{{{}} tor}}{\nu}}.
\end{equation}
Above the break frequency the decay of the emitting electrons 
is high enough that the actual volume of plasma that
contributes to the 
observed flux is smaller than that defined by the
extent of the excitation zone through the parameters
$R$ and $H$.
Because of this, the slope of the ERC spectrum
steepens by $1/2$ above 
the break frequency due to the relations $t^E_{\nu}\propto\nu^{-1/2}$
and $F^E_{{{}} \nu}\propto j^E_{{{}} \nu}t^E_{{{}} \nu}$ 
\citep[see][]{mar85}.

\subsection{Time delays}
As far as the time delays are concerned,
the ERC flares at different frequencies 
are affected by the geometry of the excitation
region and by electron energy stratification
in the same manner as synchrotron and SSC
emission.
However, certain aspects of the ERC emission variability
are markedly different.
Below, we describe the general features of the ERC flares
and the unique characteristics that distinguish them
from synchrotron and SSC flares.

The ERC light curves for the viewing angle $\theta_{obs}=0^\circ$ are
presented in Fig.~\ref{melc.0}  for $\gamma_{min}=100$ and Fig.~\ref{melc.0a} for $\gamma_{min}=10$.
The frequencies were chosen around the break frequency
at the crossing 
time so that the resulting profiles are roughly
symmetric.
At higher frequencies the profile of a flare is expected to have a flat top,
which is evident in the light curve at frequency  $\nu=2.5\times10^{18}\,\mbox{Hz}$
(dot-dashed curves).
The decay time of scattering electrons that provide the dominant contribution 
to the observed ERC flux is smaller than the apparent crossing time.
The light curves at lower frequency,  $\nu=4\times10^{17}\,\mbox{Hz}$, are symmetric because the decay time
of electrons matches the apparent crossing time (dashed curves).
All light curves peak at the crossing time in the case of $\gamma_{min}=10$.
This is to be expected since the seed emission from the torus is constant during the flare,
unlike the seed emission in the SSC model.
However, in the case of $\gamma_{min}=100$ the light curve at  $\nu=4\times10^{17}\,\mbox{Hz}$
peaks after the crossing time.
It is also evident that the spectral index of the ERC emission is
positive at the beginning of  the flare when  $\gamma_{min}=100$.
Positive values of the spectral index indicate that the light curves
are observed at frequencies below the turn-over frequency, which depends on $\gamma_{min}$.

This unusual time delay of the ERC light curve can be
understood as follows. 
The frequencies at which this delay can be observed
are below the 
turn-over frequency at the crossing time.
This means that the optimum Lorentz factors of
scattering electrons
that could provide the dominant contribution to the
observed flux
at these frequencies are below the minimum Lorentz
factor of the injected
electrons.
As the flare progresses, the minimum Lorentz factor of
the evolving electrons 
will eventually drop to the optimum values for the
emission at frequencies
below the turn-over frequency.
However, if the initial $\gamma_{{{}} min}$ is high enough, the
optimum value
might only be reached after the crossing time.
In this case, the ERC emission at the frequency that
corresponds
to this optimum value will continue to grow even after
time $t_{{{}} ac}$
when the shock front exits the excitation region and
the acceleration of electrons stops.
This phenomenon should not affect the SSC flares in
the same fashion 
since 
seed photons from a broad range of frequencies
contribute equally to the SSC emission at a given
frequency of observation.
The frequencies
at which electrons with Lorentz factor $\gamma_{min}$
emit synchrotron radiation are generally 
lower than the synchrotron 
self-absorption frequency for realistic parameters.
Therefore, synchrotron flares should not be expected to exhibit this effect, either.

The ERC light curves for the viewing angle $\theta_{obs}=90^\circ$ are
presented in Fig.~\ref{melc.90}  for $\gamma_{min}=100$ and Fig.~\ref{melc.90a} for $\gamma_{min}=10$.
At a viewing angle of $90^{\circ}$
the light curves at higher frequencies, defined by $t^E_{\nu}\lesssim{}t_{ac}$, peak at $\sim{}t_{{{}}
ac}/2$ as a result of 1) rapid decay of electrons that dominate the observed emission at higher frequencies
and 2) the circular geometry of the source along the line of sight.
At lower frequencies defined by $t^E_{\nu}\gtrsim{}t_{ac}$,
the maximum is closer to $t_{ac}$ since this is when the emission
fills the entire volume of the source.
The mechanism that causes extra delay of the ERC emission, which was described above,
affects ERC light curves at any viewing angle, including $\theta_{obs}=90^\circ$. 
However, when the viewing angle  $\sim90^{\circ}$
the effect is not as obvious since the delays
due to the geometrical shape of the source
and energy stratification are equally important.

In the calculations discussed here 
the parameters have been selected such that
inverse Compton energy losses
dominate over synchrotron losses,  $u_{rad}\approx30u_B$.
This, in particular, means that the decay time of synchrotron emission
is shorter than that calculated from synchrotron losses alone by a factor  $\sim{}u_{rad}/u_B$.
In Paper~I we neglected external emission; 
the break frequency of the synchrotron spectrum was at $\sim10^{12}\,\mbox{Hz}$
while the synchrotron self-absorption frequency $\sim10^{10}\,\mbox{Hz}$.
The dominance of ERC losses in the present calculations
shifts the break frequency of the synchrotron spectrum to $\sim10^9\,\mbox{Hz}$,
which is less than the self-absorption frequency.
Therefore, the synchrotron light curves
at all frequencies of interest originate from a volume of
the source that is limited by frequency stratification.
These light curves are expected to peak at $\sim{}t_{ac}/2$
when $\theta_{obs}\sim90^{\circ}$.
In contrast, when $\theta_{obs}=0^{\circ}$
the synchrotron flares at high
frequencies (at which $t_\nu < t_{\rm ac}$) are characterized by a quick rise,
flat top, and equally rapid decay after the crossing time.

Comparing this behavior with that of the ERC light curves,
one can assert that there must be a delay of about half 
the crossing time between the synchrotron flares and 
the ERC emission in the soft X-ray band (frequencies such that $t^E_{\nu}>t_{ac}$)
for viewing angle $\theta_{obs}\sim90^{\circ}$.
The synchrotron flare should 
 cease sharply before the peak of the ERC emission is reached
if the viewing angle $\theta_{obs}=0^{\circ}$.
This implies that the parameters of the emission region and
the external seed photon field are such that the break
frequency of the synchrotron spectrum is below the 
frequency of observation (which can be verified by observing a rather
steep synchrotron spectrum, with slope between $-1$ and $-1.5$).
If this is not the case, smaller delays must be expected.

It should be noted that the presented results are independent of the size
of the torus as long as the location of the emitting plasma in the jet
$z_p$ is adjusted by the same factor as the size of the torus $r_{tor}$. 
This ensures that the angles ${\theta'}_{min}$ and ${\theta'}_{max}$
are the same, which results in the same field of seed photons if 
the dust temperature is the same. 
On the other hand, if one keeps $z_p$ constant then 
adoption of a larger torus
results in amplification of the external seed photons
and, consequently, more rapid decay of scattering electrons.

\section{Discussion}
The study that we have conducted allows one to
distinguish 
between different emission mechanisms by means of 
the constraints placed 
on the magnitude of 
time delays
between flares and shapes of the light curves at different frequencies.
At viewing angle $\theta_{{{}} obs}=90^{\circ}$
in the rest frame of the emitting plasma,
synchrotron and inverse Compton (both SSC and ERC)
flares 
exhibit similar behavior in  terms of time delays.
The maximum time delay cannot exceed half the crossing
time $t_{{{}} ac}$,
which can crudely be equated with the duration of the
flare
when the light curves are symmetric.
For this and other viewing angles, 
the flares must be  asymmetric,
with a quick rising part and long quasi-exponential decay
if observed at  frequencies significantly below
the break frequency $\nu_b$, whose value is
different for each emission mechanism.
The flares observed at or above the break frequency
reflect the geometry of the source along the line 
of sight and are symmetric in our simulations because of 
the cylindrical geometry of the excitation volume.

Observing flares at zero viewing angle 
allows one to
distinguish 
more clearly
between emission mechanisms.
The higher frequency synchrotron light curves should peak at exactly
the
crossing time, whereas the SSC and ERC flares in the
X-ray band
have an extra time delay and, thus, peak after the
crossing time.
For this viewing angle, the flares are symmetric if
observed at or near the break frequency,
whereas the flares will have a flat top at higher frequencies.
In the case of the SSC emission, the delay is due to
the
light travel time of the synchrotron seed photons.
The ERC light curve might show similar delay
due to the decay of the minimum Lorentz factor of
electrons
if the initial value is large enough
($\gamma_{{{}} min}=100$ at $t=0$ proved sufficient
in our simulations).
If the initial $\gamma_{{{}} min}\ll100$, no extra time
delay of the ERC 
emission is possible.
In addition, if the ERC emission exhibits an extra
delay
compared to the synchrotron variability at lower
frequencies,
the spectral index of the ERC emission must be
positive
for a significant fraction of the duration of the
flare.
This is because this extra delay is only possible
for the ERC light curves at frequencies below the
initial turn-over
frequency of the SED.
On the other hand, the spectral index of the SSC
emission
will be negative even at the beginning of the SSC
flare
because the turn-over frequency of the SSC spectrum
is determined by the synchrotron self-absorption
frequency,
which is typically much lower than the characteristic
frequency of dust emission, $\nu_{{{}} tor}$.

Similar behavior of the ERC flares has been found
by \citet{sik01}. 
These authors explained the delays that they found  by appealing to
the gradient in the magnetic field along the jet,
which leads to a decrease in the critical frequency of 
synchrotron emission.
In their model the flare is produced by a plasma in a geometrically thin shell 
energized by forward/reverse shocks produced following the collision
between portions of the jet propagating with different relativistic speeds.
In contrast with our model, these authors assume that the injection of
relativistic electrons is uniform throughout the shell.
They ascribe the time delays between ERC and synchrotron flares to
gradients in magnetic field.
Their modeling is thus suitable to more prolonged flares than the ones
considered in this paper.

\begin{figure}
%\epsscale{1.0}
\plotone{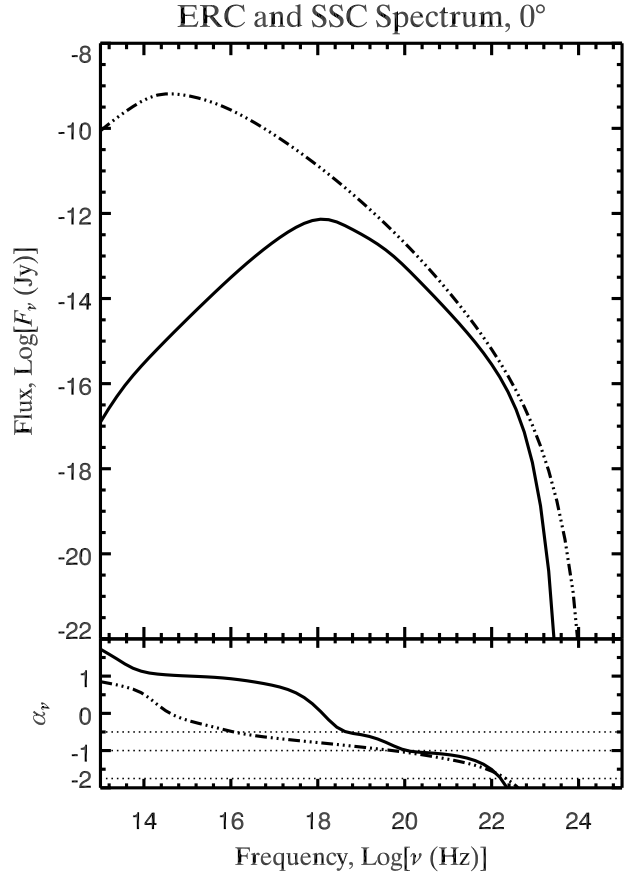}
\caption{
Same as Fig.~\ref{mesp3.0} except that the parameters
are set such 
that synchrotron self-Compton emission (triple-dot-dashed curve) dominates over
ERC radiation (solid curve).
Here the torus is assumed to be smaller and colder  
($T=800\,\mbox{K}$, $r_{{{}}tor}=7.0\,\mbox{pc}$)
while the emitting plasma is located at $z_{{{}}
p}=15\,\mbox{pc}$. 
The plasma input parameters are the same as in
Fig.~\ref{mesp.0}.
\label{mesp3.0.less}}
\end{figure}

The results reported in Paper~I are based on
the assumption
that the energy losses of electrons are dominated by
synchrotron
emission and that ERC emission is negligible
compared to
the SSC flux levels at the frequencies of interest.
These assumptions are valid when
the properties
of the 
sources of external emission as well as the
location of the emitting plasma in the jet
are as in Fig.~\ref{mesp3.0.less}.
Decreasing the size of the torus and/or placing the 
emitting plasma farther down the jet
substantially decreases the Doppler boosting
even if $\Gamma'_p$ remains the same.
For the parameters used in Fig.~\ref{mesp3.0.less}
the ERC losses of electrons are only a fraction of the
synchrotron losses.
Also, the SSC flux exceeds the ERC flux by a factor of
at least a few at
all frequencies
and by several orders of magnitude near
$10^{16}\,\mbox{Hz}$ (in the plasma rest frame),
at which the SSC flare has been shown in Paper~I
to have an extra delay due to the travel time of the seed photons.

\begin{figure}
%\epsscale{1.0}
\plotone{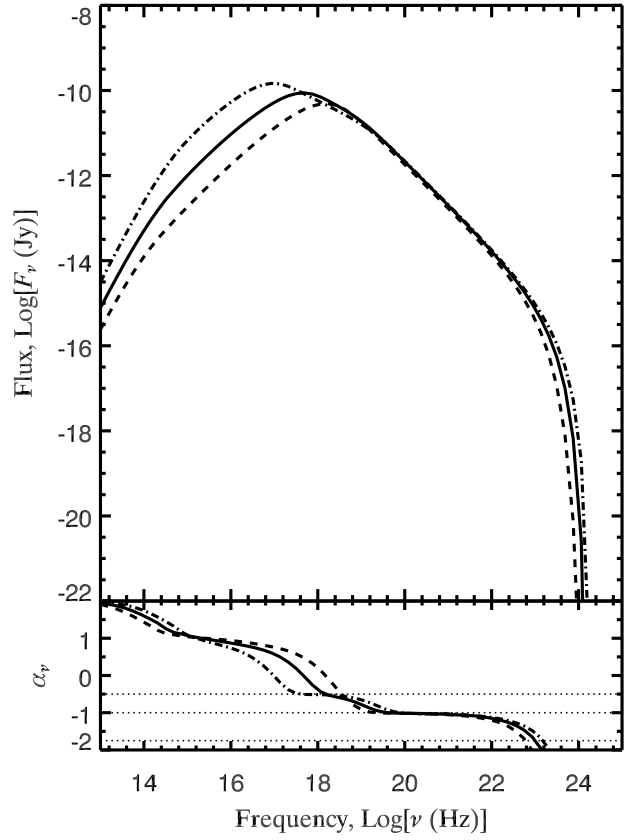}
\caption{
The dependence of the ERC spectrum on the bulk Lorentz
factor of the 
emitting plasma (dot-dashed, solid, and dashed curves
for $\Gamma'_p=5, 
10, 20$, respectively) when ERC losses of
electrons provide the 
dominant energy loss mechanism. The input parameters
are the same as in 
Fig.~\ref{mesp.0} except for $\Gamma'_p$.
\label{mesp3.0.g}}
\end{figure}

The results of simulations in the plasma rest
frame  can be transformed to the observer's frame
according to the 
following expressions: $F^{*}_{\nu_{*}}(t^{*}_{{{}} obs})
=
\delta_{{{}} obs}^3(D_0/D^{*})^2
F^E_{{{}} \nu}(t_{{{}} obs})$,
$\nu_{*}=\delta_{{{}} obs}\nu/(1+z)$, and
$t^{*}_{{{}} obs}=(1+z)t_{{{}} obs}/\delta_{{{}}
obs}$,
where the asterisk denotes quantities in the
observer's frame, $\delta_{{{}} obs}$ is the Doppler
factor determined by 
$\Gamma'_p$ and the viewing angle, and
$z$ and $D^{*}$ are the redshift and distance to
the blazar
($D_0$ is the reference distance used in the
simulations).
This implies a particular dependence of the observed
quantities
on the bulk Lorentz factor of the emitting plasma.
However, in the case where the energy losses of
electrons are dominated 
by ERC emission, this dependence will be different
due to energy stratification of the emitting volume.
Indeed, if the frequency of observation is above the
break frequency,
the size of the actual emitting volume that
contributes to the observed 
flux is smaller than the size of the excitation region
caused by the 
shock collision. In this case the thickness of the
emitting volume is determined by the decay time of the
ERC emission:
\begin{equation}
t^{{{}} E}_{\nu}\propto\frac{\sqrt{\delta_{{{}}
max}}}{u_{{{}} rad}}\propto
{\Gamma'_p}{\delta^{-2.5}_{{{}} max}}.
\end{equation}
This expression follows from the definition of the
decay time plus Eq.~(\ref{uradapprox}).
By applying the Doppler boosting formulas to the
external emission,
one can show that the ERC emission coefficient $j^{{{}}
E}_{\nu}\propto\delta^{3+(s-1)/2}_{{{}}
max}{\Gamma'}_{{{}} p}^{-1}$,
where $-s$ is the slope of the injected power-law
distribution of electron
energies. By combining these two formulas, one finds
for the ERC flux in 
the plasma rest frame
$F^E_{{{}} \nu}\propto{}j^{{{}} E}_{\nu}t^{{{}}
E}_{\nu}\propto\delta^{(s-2)/2}_{{{}} max}$.
For $s=2$ the ERC flux in the plasma rest frame does not depend on the bulk
speed of the 
emitting plasma despite relativistic boosting of the incident emission, as one
can see in 
Fig.~\ref{mesp3.0.g}.  Similar reasoning applied to
the SSC decay time 
gives $t^{{{}} C}_{{{}} \nu}\propto\Gamma'_p\delta^{-3}_{{{}} max}$.
The synchrotron emission that provides the seed
photons for the SSC 
radiation is affected in a similar way, which results
in $F^C_{{{}} \nu}\propto{}j^C_{{{}} \nu}t^C_{{{}}
\nu}\propto{\Gamma'}^2_{{{}} p}\delta^{-6}_{{{}} max}$
for the SSC flux above the break frequency, in the rest frame of the plasma.
These modifications are due to internal structure of the emitting medium
and, thus, cannot be reproduced by the homogeneous models such as \citet{der97,sik01}.
The results reported in \citet{der97}, including the boosting formulas
that give the dependence of observed flux on the Doppler factor of the emitting medium, 
should only apply to stationary emission or 
in a limited way to the flares observed at or below the break frequency.
The break in the electron energy distribution used in these studies can only occur
in a homogeneous medium with uniform and continuous injection of electrons,
and is not applicable to flares generated by acceleration of
electrons at shocks
or any other type of front.

\section{Conclusions}
To expand the realm of the calculations of Paper~I, we have
conducted
simulations of the external Compton emission
generated by collisions of shocks in a relativistic
jet in a blazar.
We have considered the physical conditions under which
the emission at high energies is dominated by either SSC
or ERC emission. If the emitting region in the jet
lies beyond the distance of the torus from the
central engine, then SSC will dominate and the results
of Paper I should be used to calculate the evolution
of the high-energy spectrum. Otherwise, ERC will be
more important because of substantial relativistic boosting 
of the external seed emission and the results of this paper are
relevant.

For the case when ERC emission dominates, we have investigated the multifrequency light curves
and determined that ERC flares at lower frequencies can incur
an extra time delay
due to the minimum Lorentz factor cut-off in the
injected distribution
of electrons. 
The simulations indicate that the spectral index of
the ERC emission
must be positive for at least half of the duration of
the flare for such a delay to occur,
which distinguishes it from a time delay of the SSC
emission, which is  characterized by
a negative spectral index at all times.
We have also found that if the energy losses of electrons
are dominated
by ERC emission, 
the dependence of the observed flux (synchrotron, SSC,
and ERC)
on the bulk speed of the emitting plasma
is different from that expected in homogeneous
models.
In particular, there is no double boosting of the ERC
emission,
while the SSC flux might even decrease 
at higher values of the bulk Lorentz factor
of the emitting plasma.

When the blazar is observed along the jet axis,
flat tops are expected in light curves 
observed at frequencies sufficiently above
the break frequency.
If the flares are symmetric, it is possible for both SSC and ERC emission
to peak in the soft X-ray band after the maximum
of the synchrotron light curve.
The value of the X-ray  spectral index during the flare
distinguishes between the SSC and ERC emission mechanism.
The latter is characterized by more shallow or even positive
spectral index.
The delayed ERC emission indicates that the minimum
Lorentz factor of the injected electrons is high enough
so that the frequency of observation is at or below the
turn-over frequency of the ERC spectrum.
Otherwise, ERC emission should peak at the same time as the synchrotron flare.

When superluminal apparent motion is observed
in VLBI images, the viewing angle cannot be zero.
In this case the behavior of the light curves is different;
it is closer to that found for $\theta_{obs}=90^{\circ}$.
Synchrotron flares that peak at IR or optical
frequencies precede the SSC and ERC flares in soft X-rays
by $\Delta{t}\sim t_{ac}/2$ owing to frequency stratification.

\acknowledgments
This material is based on work supported by NASA
grants NAG5-13074 and NNG04GO85G, as well as National
Science Foundation grant AST-0406865.

\clearpage 

\clearpage 

\clearpage 

\clearpage

\clearpage

\clearpage

\clearpage

\clearpage

\clearpage

\clearpage

\clearpage

\end{document}